\documentstyle[pra,aps]{revtex}
\begin{document}
\draft
\title
{
Quantum copying: Beyond the no-cloning theorem
}
\author{
V. Bu\v{z}ek$^{1,2}$ and
M. Hillery$^{1}$,
}
\address{$^{1}$
Department of Physics and Astronomy, Hunter College of the City University
of New York, \\
695 Park Avenue, New York, NY 10021, USA\newline
$^{2}$ Institute of Physics, Slovak Academy of Sciences, 
D\'{u}bravsk\'{a} cesta 9, 842 28 Bratislava, Slovakia
}

\date{\today}
\maketitle
\begin{abstract}
We analyze a possibility of copying ($\equiv$
cloning) of arbitrary states of quantum-mechanical
spin-1/2 system. We show that	there exists a ``universal
quantum-copying machine'' (i.e. transformation)
which approximately copies quantum-mechanical states
such that the quality of its output does not depend
on the input. We also examine a machine
which combines a unitary transformation and a selective
measurement to produce good copies of states in the neighborhood
of a particular state. We discuss the problem of measurement
of the output states.

\end{abstract}
\narrowtext
\twocolumn

\section{INTRODUCTION}
Suppose we have a quantum state $|s\rangle$ of a particular system
which we would like to copy.  For simplicity, assume that the state
space of our system is two dimensional like that of a spin-1/2 particle,
the polarizations of a photon, or a two level atom.  We shall denote the
basis elements by $|0\rangle_{a}$ and $|1\rangle_{a}$, where the subscript
$a$ is used to indicate that this is the original system (which we shall
often refer to as a mode with the polarization example in mind) which is
to be copied.  The state $|s\rangle_{a}$ is some linear combination of
$|0\rangle_{a}$ and $|1\rangle_{a}$.  We now want to feed $|s\rangle_{a}$
into a device, which we call a quantum copying machine, which at its output
will give us $|s\rangle_{a}$ back and, in addition, a copy, i.\ e.\ a system
identical to the one we put in which is also in the quantum state
$|s\rangle$.
Thus, we put in one system and get out two, both of which are in the same
quantum state as the one which was fed into the input.

Let us now make this quantitative.  The {\em ideal} copying process is
described by the transformation
$$
| s\rangle_{a}| Q\rangle_{x} \longrightarrow | s\rangle_{a}| s\rangle_{b}
| \tilde{Q}\rangle_{x},
\eqno(1.1)
$$
where $| s\rangle_{a}$ is the {\em in}-state of the original mode,
$| Q\rangle_{x}$ is the {\em in}-state of the copying device. The
{\em in}-state of the copy mode ({\em b}) is not specified in the
transformation	(1.1). In our discussion there
is no need to specify this state, though in
real physical processes  this state can be assumed to be
$| 0\rangle_{b}$ (like a blank paper in a copying machine).
The whole idea of quantum copying is to produce
at the output of the copying machine two identical states $| s\rangle_{a}$ and
$| s\rangle_{b}$ in the modes {\em a} and {\em b}, respectively. The
final state of the copying machine is described by the vector
$|\tilde{Q}\rangle_{x}$.

First one has to ask a question: ``{\em Does quantum mechanics allow
the transformation (1.1) for an arbitrary input state $| s\rangle_{a}$ ?}''.
Wootters and Zurek [1]	have answered this question. The answer is simple:
``{\em No}''. That is, quantum-mechanical states cannot be cloned
(therefore the
no-cloning theorem). To be more specific, the Wootters-Zurek no-cloning
theorem tells us that quantum states cannot be cloned ideally for an
{\em arbitrary} original {\em in}-state. This result has recently been
extended to mixed states by  Barnum, Caves, Fuchs, Jozsa, and Schumacher
[2].  They have shown
that if a particle in an arbitrary mixed state is sent into a device
and two particles emerge, it is impossible for the two reduced
density matrixes of the two-particle state to be identical to the
input density matrix.  Nevertheless, it is still an open
question  how well one can copy
quantum states, i.e. when ideal copies are not available how close the
copy state ({\em out}-state in the mode {\em b}) can be to the original state
(i.e. $| s\rangle_{a}$). The other question to answer is
what happens to the original state $| s\rangle_{a}$ after the copying.
In the present paper we will illuminate these questions. In addition
we will discuss how to measure the {\em out}-states
after the copying procedure.

Why is quantum copying of interest?  With the advent of quantum
communication, e.\ g.\ quantum cryptography, and quantum computing,
understanding the limits of the manipulations we can perform on
quantum information becomes important.	The no-cloning theorem is
one such limit.  It tells us that arbitrary quantum information
cannot be copied exactly.  On the other hand, we may be interested
only in copying a restricted set of quantum bits, or qubits, approximately.
Such a copy would allow us to gain some, but no all, information about
the original.  We would like to find out what we can do under these
less restrictive conditions.  We shall examine a number of possibilities.
We shall first study a quantum copying machine of the type proposed by
Wootters and Zurek in their proof of the no-cloning theorem.  This
machine has the property that the quality of the copy it makes
depends on the input state.  We shall next consider a copying machine
for which this is not true, i.\ e.\ the quality of the copy is the same
for all input states.  We shall also look at a machine which is designed
to copy well only a restricted set of input states.  In particular, it
copies states which are in the neighborhood of a particular state well,
but copies states which are far from this state poorly.  It would be
reasonable to use a machine of this type if we are dealing with qubits
which are near a given state.
Finally we examine briefly  the entanglement of the copy
and the original.  Because of this entanglement any measurement we
perform on the copy will change the state of the original.  We show
how nonselective measurements can be used to minimize this problem.

The paper is organized as follows. In Section II we briefly describe
the Wootters-Zurek copying procedure and we analyze quantum-statistical
properties of the copied states. We introduce measures (Hilbert-Schmidt
norms) on Hilbert space
which allow us to quantify how ``good'' the copying procedure is.
Section III is devoted to a description of the universal quantum copying
which is input-state independent. That is we will describe a copying
transformation for which the Hilbert-Schmidt norms under consideration
are input-state independent. In Section IV we will analyze a measurement
procedure by means of which one of the output states  can be measured
so the information about the second state can be obtained under the
condition that it is least perturbed by the measurement procedure.
In Section V we will discuss some specific quantum-copying transformations
which produce very good copies in the neighborhood of specific	input
states. We finish the paper with conclusions.

\section{Wootters-Zurek quantum copying machine and non-cloning theorem}

In their paper [1] Wootters and Zurek analyzed the copying process
defined by the transformation relation on basis vectors $| 0\rangle_{a}$
and $| 1\rangle_{a}$:
$$
| 0\rangle_{a}| Q\rangle_{x} \longrightarrow | 0\rangle_{a}| 0\rangle_{b}
| Q_0\rangle_{x};
\eqno(2.1a)
$$
$$
| 1\rangle_{a}| Q\rangle_{x} \longrightarrow | 1\rangle_{a}| 1\rangle_{b}
| Q_1\rangle_{x}.
\eqno(2.1b)
$$

From the unitarity of the transformation process (2.1) and the orthonormality
of the basis states $| 0\rangle_{a}$ and $| 1\rangle_{a}$  it follows that
the copying-machine states $| Q_0\rangle_{x}$ and $| Q_1\rangle_{x}$
are normalized to unity, provided that $_x\langle Q|Q\rangle_x=1$, i.e.
we can assume that
$$
_x\langle Q|Q\rangle_x=\,
_x\langle Q_0|Q_0\rangle_x=\,
_x\langle Q_1|Q_1\rangle_x =1.
\eqno(2.2)
$$
The Wootters-Zurek (WZ) quantum copying machine (QCM) is defined in such a
way that the basis vectors $| 0\rangle_{a}$ and $| 1\rangle_{a}$ are
copied ($\equiv$ cloned) ideally, that is, for these
states the relation (1.1) is fulfilled. We note that the Wootters-Zurek
copying machine is input-state dependent. Following Wootters
and Zurek we check how the pure superposition state  $| s\rangle_{a}$
(the so-called $SU(2)$ coherent state [3])
defined as
$$
| s\rangle_{a} =\alpha | 0\rangle_{a} +\beta | 1\rangle_{a}
\eqno(2.3)
$$
is copied by the copying machine described by the transformation relation
(2.1). For simplicity we will assume in what follows that probability
amplitudes $\alpha$ and $\beta$ are real and  $\alpha^2+\beta^2=1$.
Throughout the paper (except the Section 5) we consider $\alpha$
and $\beta$ to be real. Our results do not depend on this assumption
and can be easily extended for complex $\alpha$ and $\beta$.
Using the transformation relation (2.1) we obtain:
$$
| s\rangle_{a} | Q\rangle_{x} \longrightarrow
\alpha| 0\rangle_{a} | 0\rangle_{b} | Q_0\rangle_{x}
+\beta | 1\rangle_{a} | 1\rangle_{1} | Q_1\rangle_{x}
\equiv | \Psi\rangle_{abx}^{(out)}.
\eqno(2.4)
$$
If it is assumed that $_x\langle Q_0| Q_1\rangle_x=0$, i.e. the two
copying-machine states $| Q_0\rangle_{x}$ and  $| Q_1\rangle_{x}$ are
orthonormal, then the reduced density operator $\hat{\rho}^{(out)}_{ab}$
describing the state of the original-copy subsystem after the copying
procedure reads:
$$
\hat{\rho}^{(out)}_{ab} = {\rm Tr}_x \left[ \hat{\rho}^{(out)}_{abx}\right]
= \alpha^2 |00\rangle\langle 00|    + \beta^2|11\rangle\langle 11|,
\eqno(2.5)
$$
where $\hat{\rho}^{(out)}_{abx}\equiv |\Psi\rangle_{abx}^{(out)}\,
_{abx}^{(out)}\langle\Psi |$ [see Eq.(2.4)] and the basis vectors
associated with  two two-level systems under consideration are defined as
$$
|00\rangle\equiv | 0\rangle_{a}| 0\rangle_{b};\qquad
|11\rangle\equiv | 1\rangle_{a}| 1\rangle_{b};
$$
$$
|01\rangle\equiv | 0\rangle_{a}| 1\rangle_{b};\qquad
|10\rangle\equiv | 1\rangle_{a}| 0\rangle_{b}.
\eqno(2.6a)
$$
We also introduce two  vectors $|\pm\rangle$
$$
|+\rangle=\frac{1}{\sqrt{2}}(|10\rangle +|01\rangle);\qquad
|-\rangle=\frac{1}{\sqrt{2}}(|10\rangle -|01\rangle).
\eqno(2.6b)
$$
which together
with $|00\rangle$ and $|11\rangle$ create an orthonormal basis.
Density operators describing quantum states of the original mode
and the copy mode after the copying procedure read
$$
\hat{\rho}^{(out)}_{a} = {\rm Tr}_b \left[ \hat{\rho}^{(out)}_{ab}\right]
= \alpha^2 |0\rangle_a\, _a\langle 0|	 + \beta^2|1\rangle_a\, _a\langle 1|,
\eqno(2.7a)
$$
$$
\hat{\rho}^{(out)}_{b} = {\rm Tr}_a \left[ \hat{\rho}^{(out)}_{ab}\right]
= \alpha^2 |0\rangle_b\, _b\langle 0|	 + \beta^2|1\rangle_b\, _b\langle 1|,
\eqno(2.7b)
$$
respectively.
From Eq.(2.7) it follows that both the original and the copy mode at the
output are in identical states (this is a good news), but the original mode
at the output is in a mixture state (all off-diagonal elements are
destroyed).

In order to judge how good the copying machine is we need a way of comparing
its output to what, ideally, its output should be.  That is, we need a way
of comparing density matrixes.	We shall use the square of the Hilbert-Schmidt
norm of the difference between two density matrixes as a measure of how close
they are to each other.  The Hilbert-Schmidt norm of an operator, $\hat{A}$,
is given by
$$
\|\hat{A}\|_{2} = [{\rm Tr}(\hat{A}^{\dagger}\hat{A})]^{1/2},
\eqno(2.8)
$$
and it has the property that for operators $\hat{A}$ and $\hat{B}$
$$
|{\rm Tr}(\hat{A}^{\dagger}\hat{B})|\leq \|\hat{A}\|_{2}\|\hat{B}\|_{2}.
\eqno(2.9)
$$
Our distance between the density matrixes $\hat{\rho}_{1}$ and
$\hat{\rho}_{2}$ is then
$$
D=\left(\|\hat{\rho}_{1}-\hat{\rho}_{2}\|_{2}\right)^{2}.
\eqno(2.10)
$$
Is this a reasonable measure?  In a two-dimensional space any observable, $A$,
which is not a multiple of the identity, is represented by an hermitian
operator, $\hat{A}$, which can be expressed as
$$
\hat{A}=\lambda_{1}\hat{P}_{1}+\lambda_{2}\hat{P}_{2},
\eqno(2.11)
$$
where $\lambda_{1}$ and $\lambda_{2}$ are real and $\hat{P}_{1}$
and $\hat{P}_{2}$ are hermitian, one-dimensional projections with
the property that $\hat{P}_{1}\hat{P}_{2} =0$.	For a given density
matrix, $\hat{\rho}$, the probability that $A$ takes the value
$\lambda_{i}$ is given by
$$
p_{i}={\rm Tr}(\hat{\rho}\hat{P}_{i}).
\eqno(2.12)
$$
We would like our notion of closeness for density matrixes to
have the property that if two density matrixes are close, then
the probability distributions generated by them for an arbitrary
observable $A$ are also close.	That is, if $\hat{\rho}_{1}$ and
$\hat{\rho}_{2}$ are close, then the probability that $A$ takes
the value $\lambda_{i}$ in the state $\hat{\rho}_{1}$, $p_{i}^{(1)}$,
should be close to the probability that $A$ takes
the value $\lambda_{i}$ in the state $\hat{\rho}_{2}$, $p_{i}^{(2)}$.
Using the property of the Hilbert-Schmidt norm given in Eq. (2.9),
we can, in fact, show that if the Hilbert-Schmidt norm of $(\hat{\rho}_{1}
-\hat{\rho}_{2})$ is small, then $p_{i}^{(1)}$ will be close to
$p_{i}^{(2)}$.	We have
$$
\left|p_{i}^{(1)}-p_{i}^{(2)}\right|
=\left|{\rm Tr}[\hat{P}_{i}(\hat{\rho}_{1}-
\hat{\rho}_{2})]\right|
$$
$$
\leq\|\hat{P}_{i}\|_{2}\|\hat{\rho}_{1}-\hat{\rho}_{2}\|_{2}
=\|\hat{\rho}_{1}-\hat{\rho}_{2}\|_{2},
\eqno(2.13)
$$
where we have used the fact that the Hilbert-Schmidt norm of a
one-dimensional  projection is one.  This shows that our proposed
definition of closeness based
on the Hilbert-Schmidt norm has the desired property and allows us to
maintain  that, in a two-dimensional space, it is a reasonable
definition to use.

Other measures of the similarity of two density matrixes have been used.
Schumacher [4] has advocated the use of fidelity which is defined as
$$
F={\rm Tr}\left(\hat{\rho}_{1}^{1/2}\hat{\rho}_{2}\hat{\rho}_{1}^{1/2}
\right)^{1/2},
\eqno(2.14)
$$
which ranges between $0$ and $1$.  A fidelity of one means two density
matrixes are equal.  This is a more satisfactory definition in general.  The
interpretation of the Hilbert-Schmidt norm in terms of probability
distributions
breaks down in infinite dimensional spaces and becomes less good in finite
dimensional spaces as the dimension increases.	Hilbert-Schmidt norms are,
on the other hand, easier to calculate than fidelities.  For our purposes,
in a two-dimensional space, the Hilbert-Schmidt norm provides a very
reasonable way to compare density matrixes.

To see how ``far'' the copying machine drives the original mode
from its initial state	we now evaluate the Hilbert-Schmidt norm, i.e.
the ``distance'' between the {\em in}- and {\em out}-density
operators of the original mode. The Hilbert-Schmidt norm is defined as
$$
D_a \equiv {\rm Tr}\left[\hat{\rho}^{(id)}_{a}-\hat{\rho}^{(out)}_{a}
\right]^2,
\eqno(2.15)
$$
where we denote the input density operator of the original mode
as $\hat{\rho}^{(id)}_{a}$ (here index {\em id} stands for the ideal).
This density operator of the state (2.3) reads
$$
\hat{\rho}^{(id)}_{a} = \alpha^2 |0\rangle_a\langle 0|
+\alpha\beta |0\rangle_a\langle 1| +\beta\alpha |1\rangle_a\langle 0|
+\beta^2 |1\rangle_a\langle 1|.
\eqno(2.16)
$$
The Hilbert-Schmidt norm of the difference between the density
operators (2.7a) and (2.16) is
$$
D_a=2\alpha^2\beta^2=2\alpha^2(1-\alpha^2),
\eqno(2.17)
$$
which clearly reflects the fact that the states $|0\rangle_a$
(i.e. $\alpha=1$) and  $|1\rangle_a$ (i.e. $\alpha=0$) are copied
perfectly, that is for these states $D_a=0$, while the pure superposition
states $ |s\rangle_a=(|0\rangle_a\pm|1\rangle_a)/\sqrt{2}$ are copied worst.
In this case $D_a=1/2$. We remind ourselves that the maximum possible value
for  the Hilbert-Schmidt norm of the difference of two density matrixes  is
equal  to two (for instance, this is the ``distance'' between two mutually
orthogonal
states $|0\rangle_a$ and $|1\rangle_a$). If we do not specify which pure
superposition state $| s\rangle_a$ is going to be copied (i.e. we do not know
{\em a priori} the value of $\alpha$)  then on average
we should expect the distance $D_a$ in the case of the Wootters-Zurek
copying machine to be
$$
\bar{D}_a=\int_0^1 d\alpha^2 D_a(\alpha^2)=\frac{1}{3}
\eqno(2.18)
$$

From Eq.(2.17) we see that the Wootters-Zurek quantum copying procedure
is {\em state-dependent}, that is for some states it operates
well (even perfectly) while
for some states it operates badly. Moreover, as it follows from Eq.(2.5)
the output modes are , in general,
highly entangled, which is not what we would expect
from a perfect copying machine for which the output density operator
$\hat{\rho}^{(id)}_{ab}$ should be expressed as
$$
\hat{\rho}^{(id)}_{ab}=\hat{\rho}^{(id)}_{a}\otimes
\hat{\rho}^{(id)}_{b},
\eqno(2.19a)
$$
where the density operators of the ideal original and the copy at the
output are described by Eq.(2.16).  The density operator
$\hat{\rho}^{(id)}_{ab}$ in the basis (2.6) reads
$$
\hat{\rho}^{(id)}_{ab} = \alpha^4 | 00\rangle\langle 00|
+\sqrt{2}\alpha^3 \beta | 00\rangle\langle +|
+\alpha^2 \beta^2 | 00\rangle\langle 11|
$$
$$
+\sqrt{2}\alpha^3 \beta | +\rangle\langle 00|
+2\alpha^2 \beta^2 | +\rangle\langle +|
+\sqrt{2}\alpha \beta^3 | +\rangle\langle 11|
$$
$$
+\alpha^2 \beta^2 | 11\rangle\langle 00|
+\sqrt{2}\alpha \beta^3 | 11\rangle\langle +|
+\beta^4 | 11\rangle\langle 11|.
\eqno(2.19b)
$$
To measure the degree of entanglement
we can either use the entropic parameter $S_{ab}$ as proposed by Barnett
and Phoenix [5], or we can use the Hilbert-Schmidt norm $D_{ab}$
measuring the ``distance'' between the actual two-mode density operator
$\hat{\rho}^{(out)}_{ab}$ and a direct product of density operators
$\hat{\rho}^{(out)}_{a}$ and $\hat{\rho}^{(out)}_{b}$, i.e.
$$
D_{ab}^{(1)}={\rm Tr}\left[\hat{\rho}^{(out)}_{ab} -
\hat{\rho}^{(out)}_{a}\otimes\hat{\rho}^{(out)}_{b}\right]^2.
\eqno(2.20)
$$
Using the explicit expressions for the density operators which appear in
Eq.(2.20) we find  the Hilbert-Schmidt norm to be
$$
D_{ab}^{(1)}=D_a D_b,
\eqno(2.21)
$$
where the single-mode norms $D_a$ ($D_b$) are given by Eq.(2.17). Analogously
we can evaluate the Hilbert-Schmidt norm for the density operators
$\hat{\rho}^{(out)}_{ab}$ and $\hat{\rho}^{(id)}_{ab}$ [see Eqs.(2.5)
and (2.19), respectively]. In this case we find
$$
D_{ab}^{(2)}={\rm Tr}\left[\hat{\rho}^{(out)}_{ab} -
\hat{\rho}^{(id)}_{ab}\right]^2= D_a + D_b.
\eqno(2.22)
$$
Note that this result and Eq. (2.21) imply that the output state
is most entangled when the performance of the copying machine is worst.
To complete the picture we evaluate the distance $D^{(3)}_{ab}$
between the ideal
output described by the density operator $\hat{\rho}^{(id)}_{ab}$
and the direct product of the single-mode density operators
$$
D_{ab}^{(3)}={\rm Tr}\left[\hat{\rho}^{(id)}_{ab} -
\hat{\rho}^{(out)}_{a}\otimes\hat{\rho}^{(out)}_{b}\right]^2=
D_a+D_b-D_{ab}^{(1)}.
\eqno(2.23)
$$
From the above equations we clearly see that the output modes are
firstly, entangled (except the cases when $\alpha^2=0$ or
$\alpha^2=1$). Secondly, the degree of entanglement quantified via
the norm $D_{ab}^{(1)}$ is initial-state dependent (i.e. it
depends on the parameter $\alpha^2$). Thirdly, there is the
following  relationship between the norms $D_{ab}^{(i)}$
$$
D_{ab}^{(1)}\leq D_{ab}^{(3)} \leq D_{ab}^{(2)}.
\eqno(2.24)
$$
For further reference we note that the input-state averaged value [see
equations Eq.(2.18) and (2.22)]
of the norm $\bar{D}_{ab}^{(2)}$ is equal to 2/3, while the input-averaged
values of $\bar{D}_{ab}^{(1)}$ and $\bar{D}_{ab}^{(3)}$ are 2/15 and
8/15, respectively.

We finish the present section on the Wootters-Zurek quantum copying
procedure with several comments.\newline
{\bf (1)} If we assume the original ({\em a}) mode to be initially
in the mixture state
$$
\hat{\rho}^{(in)}_{a}=\alpha^2| 0\rangle_a \, _a\langle 0|
+\beta^2 |1\rangle_a \, _a\langle 1|,
\eqno(2.25)
$$
then the output density operator describing the modes {\em a}
and {\em b} after the copying is given by the same relation
(2.5) as in the case of the pure input state. This means that if the
input {\em a} mode is described by the density operator (2.25),
then the input and the output density operators in the mode {\em a}
are equal, i.e.  $D_a=0$. Nevertheless, the distance $D_{ab}^{(1)}$
reflecting a degree of entanglement between the output modes
has the value  equal to $4\alpha^4\beta^4$, i.e. is the same as for the
pure input state (2.3). This simply reflects the fact, that the
Wootters-Zurek quantum copying	machine produces a strong entanglement
between output modes  even in the case when ``classical'' states
(mixtures) are copied.

{\bf (2)}
It is interesting to note that the Wootters-Zurek QCM preserves the initial
mean value of the operator $\hat{\sigma}_z=(|1\rangle\langle 1|
-|0\rangle\langle 0|)/2$ while it completely destroys any information
about the initial mean value of the operator $\hat{\sigma}_x=
(|1\rangle\langle 0|+|0\rangle\langle 1|)/2$, that is
$$
\langle\hat{\sigma}_z \rangle_a^{(in)}=
\langle\hat{\sigma}_z \rangle_a^{(out)}=
-\frac{1}{2}\cos2\phi,
\eqno(2.26a)
$$
where we have used the parameterization $\alpha=\cos\phi$
and $\beta=\sin\phi$. On the other hand
$$
\langle\hat{\sigma}_x \rangle_a^{(in)}= \frac{1}{2}\sin2\phi,
\eqno(2.26b)
$$
while $ \langle\hat{\sigma}_x \rangle_a^{(out)}=0 $ irrespective
of the initial state of the original mode {\em a}. This means that
whatever the value of the initial variance
$\langle\left(\Delta\hat{\sigma}_x\right)^2 \rangle_a^{(in)} $
its output value is equal to 1/4.
These observations suggest that the Wootters-Zurek QCM is ``designed''
in such a way that the mean value of the operator ($\hat{\sigma}_z$)
is preserved by the copying procedure, while information associated
with  other
mean values is totally destroyed. This in turn suggests that one can think
about designing a copying machine associated with certain observation levels
[6].
\newline
{\bf (3)} Finally we briefly note that the von Neumann entropy [7] can also
be utilized to describe the ``quality'' of the
original and copy modes. The von Neumann entropy of a quantum-mechanical
system described by the density operator $\hat{\rho}$ is defined as
$$
S=-k_{\rm B} {\rm Tr}\left[\hat{\rho}\ln\hat{\rho}\right].
\eqno(2.27)
$$
Due to the Araki-Lieb theorem [8] and the fact that we consider
a conservative system for which the entropy of the complete system is
constant, and  is equal to zero  providing both modes
{\em a} and {\em b} are initially in  pure states,
the entropy of the QCM ($S_x$) after the copying
is equal to the entropy ($S_{ab}$) of the original-copy
subsystem described by the density operator $\hat{\rho}_{ab}^{(out)}$.
Besides this general property we find that
$$
S_{ab} = S_{a} =S_{b} = -k_{\rm B}\left[\alpha^2\ln\alpha^2
+\beta^2\ln\beta^2\right],
\eqno(2.28)
$$
irrespective of whether the original mode has been initially
prepared in the pure state (2.3) or a corresponding statistical mixture
described by the density operator (2.25).  If the output state were
a product of states in the $a$ and $b$ modes we would have $S_{ab}
=S_{a}+S_{b}$.	The fact that the entropy is smaller than this shows
that the modes are correlated, i.\ e.\ entangled.

\section{Input-state-independent quantum copying machine}
The Wootters-Zurek QCM suffers one significant disadvantage - its operation
depends on the state of the original input. That is, the states
$|0\rangle$ and $|1\rangle$ are copied perfectly, but the superposition states
$(|0\rangle\pm|1\rangle)/\sqrt{2}$
are essentially destroyed by this particular
copying machine in the sense that information about quantum coherences
(off-diagonal elements of the density operator in a considered basis)
is eliminated.

In what follows we describe a copying process which is {\em input-state
independent}. When using this
``universal'' quantum copying machine
(UQCM)	superposition states (2.3) are copied
 equally well for {\em any} value of $\alpha$ in a sense that
the distances $D_a={\rm Tr}[\hat{\rho}_a^{(out)}-\hat{\rho}_a^{(id)}]^2$
and $D_{ab}={\rm Tr}[\hat{\rho}_{ab}^{(out)}-\hat{\rho}_{ab}^{(id)}]^2$
do not depend on the parameter $\alpha$. In addition to this we
design the UQCM in such way that both $D_a$ and $D_{ab}$ take minimal
values.

The most general quantum-copying transformation rules for pure states
on a two-dimensional space can be written as
 $$
| 0\rangle_{a}| Q\rangle_{x} \longrightarrow \sum_{k,l=0}^{1}
| k\rangle_{a}| l\rangle_{b}| Q_{kl}\rangle_{x};
\eqno(3.1a)
$$
$$
| 1\rangle_{a}| Q\rangle_{x} \longrightarrow \sum_{m,n=0}^{1}
| m\rangle_{a}| n\rangle_{b}| Q_{mn}\rangle_{x},
\eqno(3.1b)
$$
where the states $| Q_{mn}\rangle_{x}$ are not necessarily orthonormal
for all possible values of $m$ and $n$. The general copying transformation
is very complex and it involves many free parameters
$_x\langle Q_{kl}| Q_{mn}\rangle_{x}$ which characterize the
copying machine.   In what follows we will concentrate
our attention on one particular copying-transformation which fulfills
our demands as described above. We propose  the following transformation
 $$
| 0\rangle_{a}| Q\rangle_{x} \longrightarrow
| 0\rangle_{a}| 0\rangle_{b}| Q_0\rangle_{x}
+\left[| 0\rangle_{a}| 1\rangle_{b}+
| 1\rangle_{a}| 0\rangle_{b}\right]| Y_0\rangle_{x};
\eqno(3.2a)
$$
 $$
| 1\rangle_{a}| Q\rangle_{x} \longrightarrow
| 1\rangle_{a}| 1\rangle_{b}| Q_1\rangle_{x}
+\left[| 0\rangle_{a}| 1\rangle_{b}+
| 1\rangle_{a}| 0\rangle_{b}\right]| Y_1\rangle_{x},
\eqno(3.2b)
$$
which is an obvious generalization of the WZ QCM. Due to the unitarity
of the transformation (3.2) the following relations  hold
$$
_x\langle Q_{i}| Q_{i}\rangle_{x} +
2\, _x\langle Y_{i}| Y_{i}\rangle_{x}=1;\qquad i=0,1
\eqno(3.3a)
$$
$$
_x\langle Y_{0}| Y_{1}\rangle_{x} =\,
_x\langle Y_{1}| Y_{0}\rangle_{x} =0.
\eqno(3.3b)
$$
There are still many free parameters to specify, therefore we will
further assume that the copying-machine state vectors
$| Y_{i}\rangle_{x}$ and  $| Q_{i}\rangle_{x}$ are mutually orthogonal:
$$
_x\langle Q_{i}| Y_{i}\rangle_{x} =0;\qquad i=0,1,
\eqno(3.4a)
$$
and that
$$
_x\langle Q_{0}| Q_{1}\rangle_{x} =0.
\eqno(3.4b)
$$
With these assumptions in mind we find the density operator
$\hat{\rho}_{ab}^{(out)}$ describing the modes {\em a} and {\em b}
after  copying of the pure superposition state (2.3) as:
$$
\hat{\rho}^{(out)}_{ab} =
\alpha^2 | 00\rangle\langle 00|\, _x\langle Q_0|Q_0\rangle_x
+\sqrt{2}\alpha \beta | 00\rangle\langle +|\, _x\langle Y_1|Q_0\rangle_x
$$
$$
+\sqrt{2}\alpha \beta | +\rangle\langle 00|\, _x\langle Q_0|Y_1\rangle_x
+\left[2\alpha^2\, _x\langle Y_0|Y_0\rangle_x\right.
$$
$$
+\left.2\beta^2\, _x\langle Y_1|Y_1\rangle_x  \right]  | +\rangle\langle +|
+\sqrt{2}\alpha \beta | +\rangle\langle 11|\, _x\langle Q_1|Y_0\rangle_x
\eqno(3.5)
$$
$$
+\sqrt{2}\alpha \beta | 11\rangle\langle +|\, _x\langle Y_0|Q_1\rangle_x
+\beta^2 | 11\rangle\langle 11|\, _x\langle Q_1|Q_1\rangle_x.
$$
The density operator describing the {\em a} mode can be obtained from
Eq.(3.5) by tracing over the mode {\em b} and it reads
$$
\hat{\rho}^{(out)}_{a} =
| 0\rangle_a\, _a\langle 0| \left[ \alpha^2 +\left(\beta^2\,
_x\langle Y_1|Y_1\rangle_x-\alpha^2 _x\langle Y_0|Y_0\rangle_x\right)\right]
$$
$$
+| 0\rangle_a\, _a\langle 1|\alpha\beta \left[
_x\langle Q_1|Y_0\rangle_x+\, _x\langle Y_1|Q_0\rangle_x \right]
$$
$$
+| 1\rangle_a\, _a\langle 0|\alpha\beta \left[\,
_x\langle Q_0|Y_1\rangle_x+\, _x\langle Y_0|Q_1\rangle_x \right]
\eqno(3.6)
$$
$$
+| 1\rangle_a\, _a\langle 1| \left[ \beta^2 +\left(\alpha^2\,
_x\langle Y_0|Y_0\rangle_x-\beta^2\, _x\langle Y_1|Y_1\rangle_x\right)\right].
$$
The density operator  $\hat{\rho}^{(out)}_{b}$ looks exactly the same. This
means that the states of the two modes {\em a} and {\em b} at the output
of the copying machine under consideration are equal to each other, but they
are not equal to the density operator of the {\em in}-state of the original
mode [compare Eqs.(3.6) and (2.16)]. This means that the original state
is distorted by the copying. To quantify the degree of this distortion
we evaluate the Hilbert-Schmidt norm (2.10) for the density operators
(3.6) and (2.16):
$$
D_a= 2\xi^2 (4\alpha^4 -4\alpha^2 +1) + 2\alpha^2 (1-\alpha^2)(\eta-1)^2,
\eqno(3.7),
$$
where we have introduced the notation
$$
_x\langle Y_0|Y_0\rangle_x =\, _x\langle Y_1|Y_1\rangle_x \equiv\xi;
\eqno(3.8a)
$$
$$
_x\langle Y_0|Q_1\rangle_x =\, _x\langle Q_0|Y_1\rangle_x =\,
_x\langle Q_1|Y_0\rangle_x =\, _x\langle Y_1|Q_0\rangle_x \equiv \eta/2,
\eqno(3.8b)
$$
with $0\leq\xi\leq 1/2$ and $0\leq \eta\leq 2\xi^{1/2}(1-2\xi)^{1/2}\leq
1/\sqrt{2}$, which follows from the Schwarz inequality.
The relation (3.8) further specifies ``properties'' of the copying machine
under consideration. So essentially we end up with two ``free'' parameters
which we will specify further.

As we said in the introduction, we are looking for a copying machine such that
all input original state are copied equally well, that is we want the
norm (3.7) to be independent of the parameter $\alpha^2$. This means that
one of the parameters $\xi$ or $\eta$ can be determined from the condition
$$
\frac{\partial}{\partial\alpha^2} D_a=0,
\eqno(3.9)
$$
where the norm $D_a$ is given by Eq.(3.7). From Eq.(3.9) we find that if
the parameters $\xi$ and $\eta$  are related as
$$
\eta=1-2\xi,
\eqno(3.10a)
$$
then the norm $D_a$ is	input-state  independent and it takes the value
$$
D_a=2\xi^2.
\eqno(3.10b)
$$
Taking into account the relations (3.10) and (3.3a) we can rewrite now
the density operators $\hat{\rho}^{(out)}_{ab}$ and
$\hat{\rho}^{(out)}_{a}$ [see Eqs.(3.5) and (3.6), respectively] as
$$
\hat{\rho}^{(out)}_{ab} =
\alpha^2 (1-2\xi)| 00\rangle\langle 00|
+\frac{\alpha \beta}{\sqrt{2}}(1-2\xi)| 00\rangle\langle +|
$$
$$
+\frac{\alpha \beta}{\sqrt{2}} (1-2\xi)| +\rangle\langle 00|
+2\xi | +\rangle\langle +|
+\frac{\alpha \beta}{\sqrt{2}} (1-2\xi)| +\rangle\langle 11|
$$
$$
+\frac{\alpha \beta}{\sqrt{2}}(1-2\xi) | 11\rangle\langle +|
+\beta^2 (1-2\xi)| 11\rangle\langle 11|
\eqno(3.11)
$$
and
$$
\hat{\rho}^{(out)}_{a} =
| 0\rangle_a\, _a\langle 0| \left[ \alpha^2 +\xi\left(\beta^2
-\alpha^2 \right)\right]
+| 0\rangle_a\, _a\langle 1|\alpha\beta (1-2\xi)
$$
$$
+| 1\rangle_a\, _a\langle 0|\alpha\beta (1-2\xi)
+| 1\rangle_a\, _a\langle 1| \left[ \beta^2 +\xi\left(\alpha^2
-\beta^2 \right)\right].
\eqno(3.12)
$$
We determine the  optimum value of the parameter $\xi$ from the
assumption that the distance (norm) between the two-mode density operators
$\hat{\rho}^{(out)}_{ab}$ and $\hat{\rho}^{(id)}_{ab}=\hat{\rho}^{(id)}_{a}
\otimes\hat{\rho}^{(id)}_{b}$ is input-state independent. That is, we solve
the equation
$$
\frac{\partial}{\partial\alpha^2} D_{ab}^{(2)}=0,
\eqno(3.13)
$$
where
$D_{ab}^{(2)}={\rm Tr}[\hat{\rho}^{(out)}_{ab} -
\hat{\rho}^{(id)}_{ab}]^2$ and the density operator  $\hat{\rho}^{(id)}_{ab}$
is given by Eq.(2.19b). The norm  $D_{ab}^{(2)}$ in this case can be
expressed as
$$
D_{ab}^{(2)}=(U_{11})^2 +2(U_{12})^2 + 2(U_{13})^2
+ (U_{22})^2 + 2(U_{23})^2 + (U_{33})^2,
\eqno(3.14a)
$$
with the elements $U_{ij}$ given by the relations
$$
U_{11}=\alpha^4-\alpha^2(1-2\xi);\qquad
U_{12}=\sqrt{2}\alpha\beta[\alpha^2-\frac{1}{2}(1-2\xi)];
$$
$$
U_{13}=\alpha^2\beta^2;\qquad
U_{22}=2\alpha^2\beta^2 -2\xi;
\eqno(3.14b)
$$
$$
U_{23}=\sqrt{2}\alpha\beta[\beta^2-\frac{1}{2}(1-2\xi)];\qquad
U_{33}=\beta^4-\beta^2(1-2\xi).
$$
Now the equation (3.13) can be solved with respect to the parameter
$\xi$, for which we find $\xi=1/6$.  For this value of $\xi$ the norm
$D_{ab}^{(2)}$ is $\alpha^2$-independent and its value is equal to
2/9.

\subsection{Some properties of the UQCM}
{\bf (1)} Firstly we point out that  the density operator
$\hat{\rho}^{(id)}_{a}$ is diagonal in the basis
$$
|\Phi_1\rangle_a =\alpha|0\rangle_a+\beta|1\rangle_a;\qquad
|\Phi_2\rangle_a =\beta|0\rangle_a-\alpha|1\rangle_a.
\eqno(3.15)
$$
In this basis we have  $\hat{\rho}^{(id)}_{a}=|\Phi_1\rangle_a\,
_a\langle \Phi_1|$. In this same basis the density operator
$\hat{\rho}^{(out)}_{a}$ reads
$$
\hat{\rho}^{(out)}_{a}=\frac{5}{6}|\Phi_1\rangle_a\, _a\langle \Phi_1|
+\frac{1}{6}|\Phi_2\rangle_a\, _a\langle \Phi_2|,
\eqno(3.16)
$$
from which it directly follows that the von\,Neumann entropy
of the mode {\em a} at the output of the copying machine is
$$
S_a=-k_{\rm B}\left[\frac{5}{6}\ln\left(\frac{5}{6}\right)
+\frac{1}{6}\ln\left(\frac{1}{6}\right)\right].
\eqno(3.17a).
$$
Analogously we can evaluate the von\,Neumann entropy of the
{\em ab} subsystem (or, which is the same, the entropy of the
QCM after the copying process) to be
$$
S_{ab}=S_{x}=-k_{\rm B}\left[\frac{1}{3}\ln\left(\frac{1}{3}\right)
+\frac{2}{3}\ln\left(\frac{2}{3}\right)\right].
\eqno(3.17b).
$$
We see that both the von\,Neumann entropy of each output mode
{\em a} and {\em b} separately, as well as the entropy of the
two-mode subsystem {\em ab} do not depend on the input pure state
of the original mode {\em a}. Moreover, from  the fact that the
entropies under consideration fulfill the relation
$$
S_{ab}< S_a + S_b
\eqno(3.18)
$$
it follows that there does not exist a basis in which the density
$\hat{\rho}^{(out)}_{ab}$ can be represented in a factorized form
$\hat{\rho}^{(out)}_{a}\otimes \hat{\rho}^{(out)}_{b}$. As we
will see later this entanglement between the two output modes
significantly affects the measurement procedure of the two modes
after the copying. To be more specific, any measurement performed
on mode {\em b} affects the state of the mode {\em a}.

{\bf (2)} Once we have found the basis in which both density
operators $\hat{\rho}^{(id)}_{a}$ and $\hat{\rho}^{(out)}_{a}$
are diagonal we can easily find the value of the fidelity parameter $F_a$
as introduced by Schumacher[4]. The fidelity parameter which we are
interested in is given by  Eq.(2.14) with $\hat\rho_1=\hat\rho_a^{(id)}$
and $\hat\rho_2=\hat\rho_a^{(out)}$.
In our case the fidelity is equal to a constant value $\sqrt{5/6}$ for
{\em all} input states states.	We can conclude that the UQCM
has that universal property to be the input state independent, that
is all pure states are copied equally well.

{\bf (3)} It is natural to ask how the copying machine under consideration
will copy an input state described by the statistical mixture
$$
\hat{\rho}^{(id)}_{a} = A|0\rangle_a\, _a\langle 0|
+B|0\rangle_a\, _a\langle 1|+ B|1\rangle_a\, _a\langle 0|
+(1-A) |1\rangle_a\, _a\langle 1|,
\eqno(3.19)
$$
where we assume for simplicity that $B$ is real. We note that from the
condition ${\rm Tr}(\hat{\rho}^{(id)}_a)^2\leq 1$ it follows that
$$
(1-2 A)^2 + 4 B^2\leq 1.
\eqno(3.20)
$$
We find the Hilbert-Schmidt norm $D_a$ of the difference
between the input state (3.19) and the corresponding output to be
$$
D_a=2\xi^2(1-2A)^2 + 2B^2(1-\eta)^2.
\eqno(3.21a)
$$
If we assume that $\eta=1-2\xi$, then the norm (3.21a) reads
$$
D_a=2\xi^2\left[(1-2A)^2 + 4B^2\right]\leq 2\xi^2,
\eqno(3.21b)
$$
which means that the UQCM discussed here copies
mixture states better than pure superposition states with the
same diagonal density matrix elements.

If we assume $\xi=0$ [in this case the UQCM is
identical to the WZQCM] then the statistical mixtures (3.19) such that
$B=0$ are copied perfectly in the sense that
the distance $D_a$ given by Eq.(3.21b)
is equal to zero. Nevertheless one has to be aware of the fact that
the two output modes are still strongly entangled which is reflected in
the fact that the norm $D_{ab}^{(2)}$ has a  value:
$$
D_{ab}^{(2)}={\rm Tr}\left[\hat{\rho}^{(out)}_{ab}
-\hat{\rho}^{(id)}_{ab}\right]^2= 4A^2(1-A)^2.
\eqno(3.22)
$$

{\bf (4)} We can compare the performance of the Wootters-Zurek
QCM and the UQCM as discussed above if we compare the averaged values
of the norms $D_a$ and $D_{ab}$ for the WZ QCM, which read respectively
$$
\bar{D}_a=1/3;\qquad \bar{D}_{ab}=2/3
$$
with the input-state independent values of these parameters
for the UQCM. We see that in the case of the UQCM
the norm $D_a$ is 6 times smaller (on average)
while $D_{ab}$ is 3 times smaller compared to
$\bar{D}$ and $\bar{D}_{ab}$, respectively.  These relations simply
reflect the ``high-quality'' performance of the UQCM. It is still an open
 question whether the UQCM is the best (on average)
QCM quantum mechanics  would allow.

{\bf (5)}
The UQCM has that interesting property that the mean values of the
operators $\hat{\sigma}_x$ and $\hat{\sigma}_z$ are scaled by copying.
It can be found that irrespective of whether the input mode is in a pure
state or a statistical mixture the following relations	hold
$$
\langle \hat{\sigma}_j\rangle^{(out)}
=(1-2\xi)\langle \hat{\sigma}_j\rangle^{(in)}; \qquad j=x,z,
\eqno(3.23)
$$
where the relation $\eta=1-2\xi$ has been taken into account. Obviously,
for $\xi=1/6$ both $\langle \hat{\sigma}_z\rangle^{(out)}$ and
$\langle \hat{\sigma}_x\rangle^{(out)}$ are scaled by the factor 2/3.
This is in contrast to the WZ QCM, when the mean value of the operator
$\hat{\sigma}_z$ is preserved in the copying process, while
$\langle \hat{\sigma}_x\rangle^{(out)}=0$ irrespective of the input state.
We note that the relations between the input and output
mean values can be taken as definitions of particular copying machines.
To be specific, one can associate the copying process
with a given observation level,
i.e. a set of observables, and impose particular conditions on input
and output values of the observables associated with the given observation
level. The relations between the input and output mean values then can
be solved with respect to those parameters which specify the
copying machine itself, i.e. the values $_x\langle Q_{kl}|Q_{mn}\rangle_x$.

{\bf (6)} The four state vectors $|Q_0\rangle_x$,
$|Q_1\rangle_x$, $|Y_0\rangle_x$, and $|Y_1\rangle_x$ in terms of which
the QCM transformation (3.2) is defined are not orthonormal. Using the
Gram-Schmidt procedure one can defined a set of four orthonormal
quantum-copying-machine basis states $|\bar{Q}_0\rangle_x$,
$|\bar{Q}_1\rangle_x$, $|\bar{Y}_0\rangle_x$, and $|\bar{Y}_1\rangle_x$.
If we assume the relations (3.3), (3.4) and (3.8)
defining the QCM under consideration
and the relation $\eta=1-2\xi$, the orthonormal states read:
$$
|\bar{Q}_0\rangle_x=\frac{|Q_0\rangle_x}{\sqrt{1-2\xi}};\qquad
|\bar{Q}_1\rangle_x=\frac{|Q_1\rangle_x}{\sqrt{1-2\xi}};
$$
$$
|\bar{Y}_0\rangle_x=\frac{2|Y_0\rangle_x-|Q_1\rangle_x}{\sqrt{6\xi-1}};\qquad
|\bar{Y}_1\rangle_x=\frac{2|Y_1\rangle_x-|Q_0\rangle_x}{\sqrt{6\xi-1}},
\eqno(3.24)
$$
from which it follows that one has to treat carefully the case of $\xi=1/6$,
i.e. when the norm $D_{ab}$ is the input-state independent.
To be more specific, under the conditions given by Eqs.(3.3), (3.4) and (3.8)
imposed on the copying-machine vectors $|Q_i\rangle_x$ and $|Y_i\rangle_x$
($i=0,1$) with $\eta=2/3$ and $\xi=1/6$ we find the following relations
$$
_x\langle Q_i|Q_i\rangle_x =2/3;\qquad
 _x\langle Y_i|Y_i\rangle_x =1/6; \qquad i=0,1
$$
$$
_x\langle Y_1|Y_0\rangle_x = \, _x\langle Q_1|Q_0\rangle_x =0;
\eqno(3.25)
$$
$$
_x\langle Y_0|Q_1\rangle_x = \, _x\langle Y_1|Q_0\rangle_x =1/3,
$$
which means that the four copying-machine
vectors are not linearly independent. They
in fact lie in a 2-dimensional sub-space of the original 4-dimensional space
of the copying machine. In this 2-dimensional sub-space the copying-machine
vectors  have the following components:
$$
|Y_0\rangle_x= (1/\sqrt{6},0);\qquad  |Y_1\rangle_x= (0,1/\sqrt{6});
$$
$$
|Q_0\rangle_x= (0,\sqrt{2/3});\qquad  |Q_1\rangle_x= (\sqrt{2/3},0).
\eqno(3.26)
$$
We see that the vectors $|Y_i\rangle_x$ can be expressed in terms of the
vectors  $|Q_i\rangle_x$:
$$
|Y_0\rangle_x=\frac{1}{2} |Q_1\rangle_x\qquad
|Y_1\rangle_x=\frac{1}{2} |Q_0\rangle_x .
\eqno(3.27)
$$
If we introduce two orthonormal basis states $|\uparrow\rangle$ and
$|\downarrow\rangle$ in the  two-dimensional state space, then we can
express the copying-machine states $ |Q_0\rangle_x$ in this basis
as
$$
|Q_0\rangle_x = \sqrt{\frac{2}{3}}|\uparrow\rangle ;
\qquad
|Q_1\rangle_x = \sqrt{\frac{2}{3}}|\downarrow\rangle.
\eqno(3.28)
$$
Consequently, the UQCM transformation (3.2) now reads
 $$
| 0\rangle_{a}| Q\rangle_{x} \longrightarrow
\sqrt{\frac{2}{3}} | 00\rangle|\uparrow\rangle
+\sqrt{\frac{1}{3}} | +\rangle|\downarrow\rangle;
\eqno(3.29a)
$$
 $$
| 1\rangle_{a}| Q\rangle_{x} \longrightarrow
\sqrt{\frac{2}{3}} | 11\rangle|\downarrow\rangle
+\sqrt{\frac{1}{3}} | +\rangle|\uparrow\rangle,
\eqno(3.29b)
$$
where the initial copying-machine state $| Q\rangle_{x}$ can be expressed
as a linear superposition of the two basis states $|\uparrow\rangle$
and $|\downarrow\rangle$.

\section{Measurement of the original and the copy state at the output of
QCM}
From our previous discussion it follows that the original and the
copy states (described by density operators $\hat{\rho}^{(out)}_{a}$
and $\hat{\rho}^{(out)}_{b}$, respectively) are highly entangled
(see discussion in Section III). This means that any measurement
performed on the mode {\em b} will significantly affect the
state of the mode {\em a}.  This could defeat the purpose of a
quantum copying machine because by measuring the copy we distort
the original.  Ideally we would like to have the copy and
original independent so that if one is measured the other is
undisturbed and available for future processing.  We need to
determine how close the copying machine of the previous section
comes to this ideal.

What we shall do is to consider the effect of an unconditioned
measurement of the $b$ mode on the state of the $a$ mode.  Define
the $b$ mode vector
$$
| s\rangle_b= u| 0\rangle_b +v| 1\rangle_b;\qquad |u|^2 +|v|^2=1,
\eqno(4.1)
$$
and the corresponding projection operator $\hat{P}_{|s\rangle_{b}}
=|s\rangle_{b}\, _{b}\langle s|$.  We start with an ensemble of copies
and originals which has been produced by the copying machine and is
described by the density matrix $\hat{\rho}_{ab}^{(out)}$.  We now
measure $\hat{P}_{|s\rangle_{b}}$ for each element of the
ensemble and, irrespective of the result, keep the resulting
two-mode state.  This results in the new ensemble
$$
\hat{\rho}_{ab}^{(meas)}=\hat{P}_{|s\rangle_{b}}\hat{\rho}_{ab}^{(out)}
\hat{P}_{|s\rangle_{b}}+\hat{Q}_{|s\rangle_{b}}
\hat{\rho}_{ab}^{(out)}\hat{Q}_{|s\rangle_{b}},
\eqno(4.2)
$$
where $\hat{Q}_{|s\rangle_{b}}=\hat{I}_{b}-\hat{P}_{|s\rangle_{b}}$
and $\hat{I}_{b}$ is the $b$ mode identity operator.  From this
we obtain the $a$ mode reduced density matrix
$$
\hat{\rho}_{a}^{(meas)}={\rm Tr}_{b}(\hat{\rho}_{ab}^{(meas)}).
\eqno(4.3)
$$
The measurement of $\hat{P}_{|s\rangle_{b}}$ can yield either $0$
or $1$.  The probability of obtaining $1$ is given by
$$
{\rm Tr}(\hat{P}_{|s\rangle_{b}}\hat{\rho}_{ab}^{(out)}
\hat{P}_{|s\rangle_{b}})=\frac{1}{6}+\frac{2}{3}|\alpha
u^{\ast}+\beta v^{\ast}|^{2}.
\eqno(4.4)
$$
It is clear from this equation that measurement of this probability
will give us information about $\alpha$ and $\beta$.  Thus by
measuring the $b$ mode we do gain information about the quantum
state of the input mode of the copying machine.

Now let us see what the effect of the $b$-mode measurement is on
the $a$ mode.  We note that
$$
\hat{\rho}_{a}^{(out)}={\rm Tr}_{b}\left[(\hat{P}_{|s\rangle_{b}}+
\hat{Q}_{|s\rangle_{b}})\hat{\rho}_{ab}^{(out)}
(\hat{P}_{|s\rangle_{b}}+\hat{Q}_{|s\rangle_{b}})\right]
  =  \hat{\rho}_{a}^{(meas)}
 \eqno(4.5)
 $$
 so that the $a$-mode density matrix after the unconditional
 measurement is the same as that before it.  This result does
 not depend on $u$ and $v$ so we can choose to measure any
 projection in the $b$ mode.

 Even though $\hat{\rho}_{a}^{(out)}$ and $\hat{\rho}_{a}^{(id)}$
 are close ($D_{a}=1/18$) they are not the same.  However,
 because of the form of $\hat{\rho}_{a}^{(out)}$ it is possible
 to recover the expectation value of any operator in the state
 $\hat{\rho}_{a}^{(id)}$ from it.  In order to show this we
 express $\hat{\rho}_{a}^{(out)}$ as
 $$
 \hat{\rho}_{a}^{(out)}=\frac{2}{3}\hat{\rho}_{a}^{(id)}
 +\frac{1}{6}\hat{I}_{a}.
 \eqno(4.6)
 $$
 This implies that the density matrixes differ in a way that
 does not depend on $\alpha$ and $\beta$.  Therefore, if
 $\hat{A}_{a}$ is an $a$-mode operator, then
 $$
 {\rm Tr}(\hat{A}_{a}\hat{\rho}_{a}^{(id)})=\frac{3}{2}
\left[{\rm Tr}(\hat{A}_{a}\hat{\rho}_{a}^{(out)})-\frac{1}{6}
 {\rm Tr}(\hat{A}_{a})\right],
 \eqno(4.7)
 $$
 where ${\rm Tr}(\hat{A}_{a})$ does not depend on $\alpha$ and
 $\beta$ and is, therefore, known.

 In summary the output from the UQCM has the following property.
 If any projection is measured in the $b$ mode the unconditioned
 $a$-mode ensemble which results is close to the ideal output
 state, i.\ e.\ the input state, and can be used to find the
 expectation value of any $a$-mode operator in the ideal output
 state.  In addition, the $b$-mode measurement provides us with
 information about the input state.

\section{Copying states in the neighborhood of given state}
Suppose that we want to build a copying
machine which will copy to a high degree
of fidelity, states in the neighborhood of a given quantum state.
In order to
get an idea of how to construct such a machine let us look at what would
happen
if the machine made perfect copies.  We shall consider the input state given
by Eq.(2.3) where $\beta $ is close to one and
$\alpha $ is small in magnitude (in this section we consider $\alpha$ and
$\beta$ to be complex numbers because this does matter to the results).
  If this state were copied perfectly we would
have $\hat{\rho}_{ab}^{(id)}$ given by Eq.(2.19b).
Under the stated conditions on $\beta$ and $\alpha$ we have roughly that
$$
 \alpha |0\rangle_a+\beta |1\rangle_a \longrightarrow \beta^{2}|11\rangle +
\alpha (|10\rangle + |01\rangle ).
\eqno(5.1)
$$
This suggests that the copying machine specified by
$$
\begin{array}{rcl}
|1\rangle_a |Q\rangle_x & \longrightarrow & |1\rangle_a|1\rangle_b
|Q_1\rangle_x  \\
|0\rangle_a |Q\rangle_x & \longrightarrow & \frac{1}{\sqrt{2}}
(|1\rangle_a|0\rangle_b +
|0\rangle_a |1\rangle_b) |Q_1\rangle_x ,
\end{array}
\eqno(5.2)
$$
where $|Q\rangle_x $ and $|Q_1\rangle_x $ are the initial and final
states of the
copying machine, respectively, would produce something like the desired
action.  Note that this machine is very different from the Wootters-Zurek
machine  in that while
one basis vector is duplicated exactly, the other is completely changed.  In
fact, the state $|0\rangle_a $ is sent into a state which has no overlap at
all  with the perfectly cloned state $|0\rangle_a|0\rangle_b $.

We shall examine the action of this copying machine, but it is
worthwhile to note
at the beginning that there is a major problem with it.  The factor
of $1/\sqrt{2}$,
which is required by unitarity, means that we do not obtain the action
indicated in  Eq.(5.1).  What we have is that
$$
\left(	\alpha |0\rangle_a+\beta |1\rangle_a  \right) |Q\rangle_x
\longrightarrow \left[\beta|11\rangle
 + \alpha|+\rangle\right]|Q_1\rangle_x
$$
$$
\equiv |\Psi_1\rangle_{ab}^{(out)}
|Q_1\rangle_x.
\eqno(5.3)
$$
We can determine how good  job this copying
machine does by looking at the difference
between what it does and what it is supposed to do, i.e. we evaluate the
Hilbert-Schmidt norm between the states described by the density operators
$\hat{\rho}^{(id)}_{ab}$ [see Eq.(2.12b)] and
$\hat{\rho}^{(out)}_{ab}$ [see Eq.(5.3)]. For the Hilbert-Schmidt norm
$D^{(2)}_{ab}$ given by Eq.(2.15) we find the explicit expression
$$
D^{(2)}_{ab}=2-(\beta + \beta^{\ast})(|\beta |^{2}+|\alpha |^{2}
\sqrt{2}).
\eqno(5.4)
$$
This can be simplified by expressing $\beta $ as $1-\delta \beta $ and using
the  normalization condition $|\beta |^{2}+|\alpha |^{2}=1$ to find the
condition
$$
\delta \beta +\delta \beta^{\ast}= |\alpha |^{2}+|\delta \beta |^{2}.
\eqno(5.5)
$$
We then have that
$$
D^{(2)}_{ab}= (3-2\sqrt{2})|\alpha|^{2}+|\delta \beta |^{2},
\eqno(5.6)
$$
where terms of order $|\alpha |^{4}$ and $|\delta \beta |^{2}|\alpha|^{2}$
have  been dropped.  Finally, we need to determine the size of
$|\delta \beta |^{2}$.
Setting $\delta \beta = re^{i\theta}$ and substituting this into Eq.(5.5) we
find that
$$
2r\cos \theta - r^{2}= |\alpha|^{2}.
\eqno(5.7)
$$
This implies that unless $\theta $ is very close to $\pi /2$, then $r$ will be
of order $|\alpha|^{2}$.  If $\delta \theta = \theta - \pi /2$ is of order
$|\alpha |$
or less, then r is of order $|\alpha |$.  In either case, the right-hand side
of Eq.(5.6) will be of order $|\alpha|^{2}$.

It is possible to do better than this in a certain sense.  Consider the
copying machine specified by
$$
\begin{array}{rcl}
|1\rangle_a |Q\rangle_x & \longrightarrow & \frac{1}{\sqrt{2}}
\left(|11\rangle |Q_1\rangle_x
+ |00\rangle |Q_0\rangle_x \right) \\
|0\rangle_a |Q\rangle_x & \longrightarrow & \frac{1}{\sqrt{2}}\left(|10\rangle +
|01\rangle \right) |Q_1\rangle_x ,
\end{array}
\eqno(5.8)
$$
where $_x\langle Q_0| Q_0\rangle_x=\, _x\langle Q_1| Q_1\rangle_x=1$
and $_x\langle Q_0| Q_1\rangle_x=0$.
With
this copying machine we find that a superposition state goes into
$$
\alpha |0\rangle_a + \beta |1\rangle_a \longrightarrow \frac{1}{\sqrt{2}}
\left[\beta
|11\rangle + \alpha (|10\rangle +|01\rangle )\right]|Q_1\rangle_x
$$
$$
+\frac{\beta}
{\sqrt{2}}|00\rangle |Q_0\rangle_x .
\eqno(5.9)
$$
We define the vector in the first term on the right-hand side of Eq.(5.9)
to be
$$
|\Psi_{2}\rangle_{ab}^{(out)} =
\beta |11\rangle + \alpha (|10\rangle + |01\rangle ).
\eqno(5.10)
$$
This vector
 is much closer to the vector on the right-hand side of Eq.(5.1)
than is $|\Psi_{1}\rangle^{(out)}_{ab}$ given by Eq.(5.3).
In fact, we find that the Hilbert-Schmidt norm between the density operator
associated state vector (5.10) and the ideally copied state given by the
density operator (2.19b) is
$$
D^{(2)}_{ab}=|\delta \beta |^{2}+2|\alpha |^{2}|\delta \beta |^{2}.
\eqno(5.11)
$$
As long as $\theta$ is not too close to $\pi /2$, the right-had side of
this equation
will be of order $|\alpha |^{4}$.  This is a considerable improvement
over what  the Wootters-Zurek copying machine can do.  There is, however,
in this case,  the problem
of the term proportional to $|00\rangle $ in Eq.(5.9).	What we can do with the
output of this copying machine is to use it to calculate the expectation
values
of any operator which annihilates this state.  That is, if $\hat{S}$ is an
operator which
has the property that $\hat{S}|00\rangle = 0$, then we can get a very good
estimate
of the expectation value of $\hat{S}$ in the state (2.19b) by calculating
the expectation of $\hat{S}$ in the state on the right-hand side of Eq.(5.9)
and
multiplying the result by 2.  In this sense the copying machine specified by
Eq.(5.8) does a good job of copying states in the neighborhood of $|1\rangle $.

Another possibility is to use a selective measurement to obtain the desired
state from that in Eq. (5.9).  If we measure the operator $\hat{P}_{00}=
|00\rangle \langle 00|$ and obtain the value $0$ the resulting two-mode
density matrix is
$$
\hat{\rho}_{ab}^{(sel)}=\frac{1}{1+|\alpha|^{2}}|\Psi_{2}\rangle_{ab}^{(out)}
\, ^{(out)}_{ab}\langle \Psi_{2}|.
\eqno(5.12)
$$
This produces the desired result because $D_{ab}^{(2)}$ is again of the
order of $|\alpha|^{4}$ as long as $\theta$ is not too close to $\pi/2$.
A nonselective measurement of any one-dimensional projection in the $b$
mode now gives us information about $\alpha$ and $\beta$ and leaves us
with the reduced $a$-mode density matrix
$$
\hat{\rho}_{a}^{(sel)}=\frac{1}{1+|\alpha|^{2}}(\hat{\rho}_{a}^{(id)}
+|\alpha|^{2}|1\rangle \langle 1|),
\eqno(5.13)
$$
which, for $|\alpha|^{2}<<1$, is close to $\hat{\rho}_{a}^{(id)}$.
Therefore, the transformation is Eq. (5.8) followed by a selective
measurement gives us a good approximation to cloning for a limited range
os states.  The copy can be measured, providing information about the
initial state, and the resulting $a$-mode density matrix is close to
that of the input.

\section{Conclusions}
The Wootters-Zurek no-cloning theorem forbids the copying of an
arbitrary quantum state.  If one does not demand that the copy
be perfect, however, possibilities emerge.  We have examined
a number of these.  A quantum copying machine closely related to
the one used by Wootters and Zurek in the proof of their
no-cloning theorem copies some states perfectly and others poorly.
That is, the quality of its output depends on the input.  A
second type of machine, which we called a universal quantum
copying machine, has the property that the quality of its output
is independent of its input.  Finally, we examined a machine
which combines a unitary transformation and a selective
measurement to produce good copies of states in the neighborhood
of a particular state.

A problem with all of these machines is that the copy and
original which appear at the output are entangled.  This
means that a measurement of one affects the other. We
found, however, that a nonselective measurement of the
one of the output modes will provide information about
the input state and not disturb the reduced density matrix
of the other mode.  Therefore, the output of these xerox
machines is useful.

There is further work to be done; we have only explored
some of the possibilities.  It would be interesting to know,
for example, what the best input-state independent quantum
copying machine is.  One can also consider machines which make
multiple copies.  Does the quality of the copies decrease as
their number increases?  These questions remain for the future.

\vspace{1.5truecm}

{\bf Acknowledgements}\newline
This work was supported   by the National Science Foundation under grants
INT 9221716 and PHY-9403601 and by
the  East-West Program of the Austrian
Academy of Sciences under the contract No. 45.367/1-IV/6a/94 of the
\"{O}sterreichisches Bundesministerium f\"{u}r Wissenschaft und Forschung.

\end{document}